\documentclass[cameraready]{Interspeech}
\title{wav2VOT: automatic estimation of voice onset time, closure duration, and burst realisation with wav2vec2}

\author[affiliation={1}, orcid=0000-0001-8143-2099]{James}{Tanner}
\author[affiliation={2}, orcid=0000-0001-7675-2370]{Morgan}{Sonderegger}
\author[affiliation={1}, orcid=0000-0001-7400-9436]{Jane}{Stuart-Smith}
\author[affiliation={3}, orcid=0000-0002-0989-1765]{Tyler}{Kendall}
\author[affiliation={4}, orcid=0009-0002-9882-8254]{Jeff}{Mielke}
\address{
    $^1$ University of Glasgow, United Kingdom \\
    $^2$ McGill University, Canada \\
    $^3$ University of Oregon, United States \\
    $^4$ North Carolina State University, United States
}

\email{james.tanner@glasgow.ac.uk, morgan.sonderegger@mcgill.ca, jane.stuart-smith@glasgow.ac.uk, tsk@uoregon.edu, jimielke@ncsu.edu}

\keywords{voice onset time, wav2vec2, stops, automatic annotation}

\newcommand{\blue}[1]{\textcolor{blue}{#1}}

\usepackage{comment}

\begin{document}

\maketitle

\begin{abstract}
While automatic tools for speech annotation are now commonplace within phonetic research pipelines, many tasks require substantial manual correction or training sets to perform accurately. Simultaneously, large speech models such as wav2vec2 have been shown to perform well at speech classification tasks, raising the question of how these models may be applied to phonetic annotation tasks. We introduce wav2VOT: a tool for the automatic estimation of voice onset time, closure duration, and burst realisation using wav2vec2. We demonstrate that wav2VOT performs comparably with current approaches on unseen datasets, and can estimate with high accuracy with fine-tuning. Analysis of wav2VOT predictions demonstrate high fidelity across stop voicing and place of articulation. These results demonstrate that large speech models are capable of producing accurate annotations, and further motivate exploration of large speech models as tools in phonetic research pipelines.
\end{abstract}

\section{Introduction}
Automatic approaches to the segmentation and annotation of speech data has become increasingly popular within phonetic and speech science research. Many analyses now utilise a pipeline of (semi-)automatic utterance- and word-level transcription and automatic time-alignment of phonemic labels to the speech signal (e.g. \cite{coats2020,tanner2020VE,brand2021}), with further downstream annotations of sub-phonemic variables depending on specific research questions. One such widely-studied phonetic variable is that of Voice Onset Time (VOT), which reflects the temporal interval between the release of a stop consonant and the onset of voicing for the subsequent vocalic segment \cite{abramson2017}. VOT has been long-shown to distinguish phonological voicing categories in stops across languages \cite{lisker1964}, and can vary as both `negative VOT' (voicing starting before the stop release) and `positive VOT` (voicing occurring following the stop release). As a result, VOT, along with other acoustic cues such as closure duration and probability of stop reduction \cite{lisker1977,port1979,cho1999}, have been intensively studied as key variables for laryngeal and voicing systems across the world's languages \cite{salmons2019}.

% - automated methods increasingly used within phon research
%     - esp movement towards 'large-scale' ling analysis, enabled by large speech corpora (SPADE, Coats, ref Liberman)
%     - typical modern speech pipeline: transcription, alignment, then some kind of acoustic analysis
%     - this acoustic analysis might also require further downstream annotation processes

% - VOT is widely-studied phon variable:
    % - follows given patterns (short/long, brief mention of neg)

While VOT is a frequently-studied acoustic variable within phonetic research, manual VOT annotation is often a labourious and resource-intensive task, placing a functional upper-limit on the amount of stop tokens that can be feasibly annotated, in turn resulting in numerous tools that have been developed to automate this process. The most widely-used application, AutoVOT \cite{autovot}, utilises a pre-trained classifier to generate VOTs from a set of user-provided audio files, and has found widespread application in phonetic and clinical studies of VOT variation \cite{chodroff2017,buz2018,sondereggerlanguage}. Although AutoVOT and other tools utilising different methods \cite{chi-yueh2011,pratosh2014,adi2016sequence,shrem2019dr,getvot} have demonstrated that VOT can be automatically estimated, the functionality of these tools also create limitations in how they may be applied for the annotation of VOT and the acoustic properties of stops. First, with the exception of DeepSegmentor \cite{adi2016sequence}, these approaches exclusively provide annotations of VOT, leaving other challenging tasks such as closure duration annotation to be performed via alternative methods \cite{sondereggerlanguage}. Second, these tools assume that the stop is realised with a distinct burst, which may not be the case given that stops frequently undergo lenition \cite{Bouavichith2013,priva2020}. Third, these existing tools often require extensive data formatting, in turn requiring proficiency in programming and data processing.

% - prev work on automatic  VOT estimation:
%     - main/most prominent = AutoVOT (briefly describe how it works)
%     - also mention the others

% - issues (with AutoVOT):
%     - 1. no annotation of closure duration: while VOT important, wide range of studies have shown the importance of CD in marking phon contrasts, yet no tool exists
%     - 2. no annotation of stop lenition/reduction: assumes all stops given contain a measureable burst section, but is often not a valid assumption (esp for connected and spontaneous speech)
%     - 3. formatting data for use can be difficult (esp for technical users)

At the same time, there is now an increasing body of research exploring the potential of applying large speech models such as wav2vec2 \cite{wav2vec,wav2vec2} to various speech annotation tasks, such as the classification of stop consonant lenition \cite{tanner2025interspeech}, variable nasality \cite{kim2024,zellou2024}, and automatic transcription \cite{patman2024,pater2025}.  It has been shown, however, that the architecture of wav2vec2 places some limits on the capacity to perform tasks such as forced alignment \cite{zhu2022,rousso2024traditioninnovationcomparisonmodern}. Together, these findings motivate further exploration of the capacity of large speech models (such as wav2vec2) for phonetic annotation tasks, including the modification of the model architecture for specific tasks.

% This, combined with growing evidence of the capacity for the wav2vec2 architecture to latently represent phonetic knowledge \cite{dieck2022,deheerkloots2024}, points to the potential for large pre-trained speech models such as wav2vec2 for phonetic annotation tasks.

% - this, combined with studies showing detailed phon knowledge, points to PTSMs being great tools for building phon annotators

% - at the same time, researchers also exploring the potential of modern large speech models, most prominently 'self-supervised' speech models like wav2vec
%     - brief description of wav2vec2:
%         - FE + transformer
%         - pre-trained on large amount of speech data (via self-supervision) and previously used for wide range of speech tasks
%     - nasalisation annotation, burst realisation
%     - transcription (Pater AutoIPA)
%     - french group work
%     - chodroff alignment

This study explores this question and presents wav2VOT: a tool for the automatic estimation of VOT, closure duration, and burst realisation utilising the wav2vec2 architecture.\footnote{Source code and models available at\\\url{https://github.com/james-tanner/wav2VOT}.} Section \ref{sec:model} discusses the technical modifications to the wav2vec2 architecture necessary for fine-grained temporal prediction. We then report two types of evaluation for the viability of wav2VOT as a practical tool for stop annotation: first, through the evaluation of the predictive performance of wav2VOT on multiple speech datasets, including the effect of finetuning the model for specific data distributions (Sec.~\ref{sec:ft}), and second comparing wav2VOT predictions directly with hand-annotated VOT and closure duration in a hypothetical phonetic study context (Sec.~\ref{sec:analysis}).

% - this work extends this logic and applies to the annotation of stop realisation: specifically VOT and closure duration

% - this paper demonstrates application of wav2vec2 for the automatic annotation of stop VOT and closure duration, incl its use across speech datasets of different langs, speech style, and recording quality

\section{Model}
\label{sec:model}

\begin{figure*}[h]
    \centering
    \includegraphics[width=0.33\linewidth]{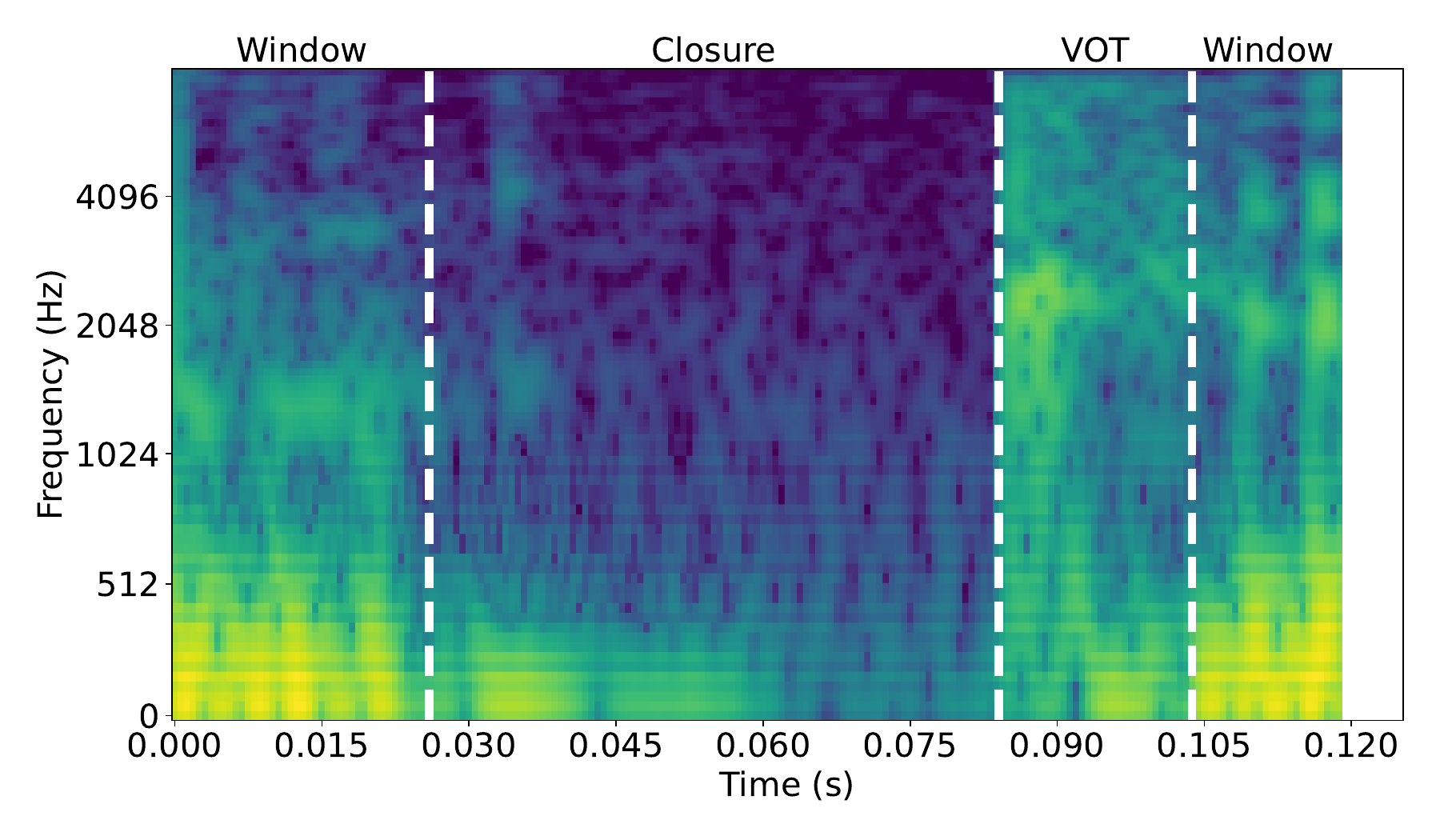}
    \includegraphics[width=0.33\linewidth]{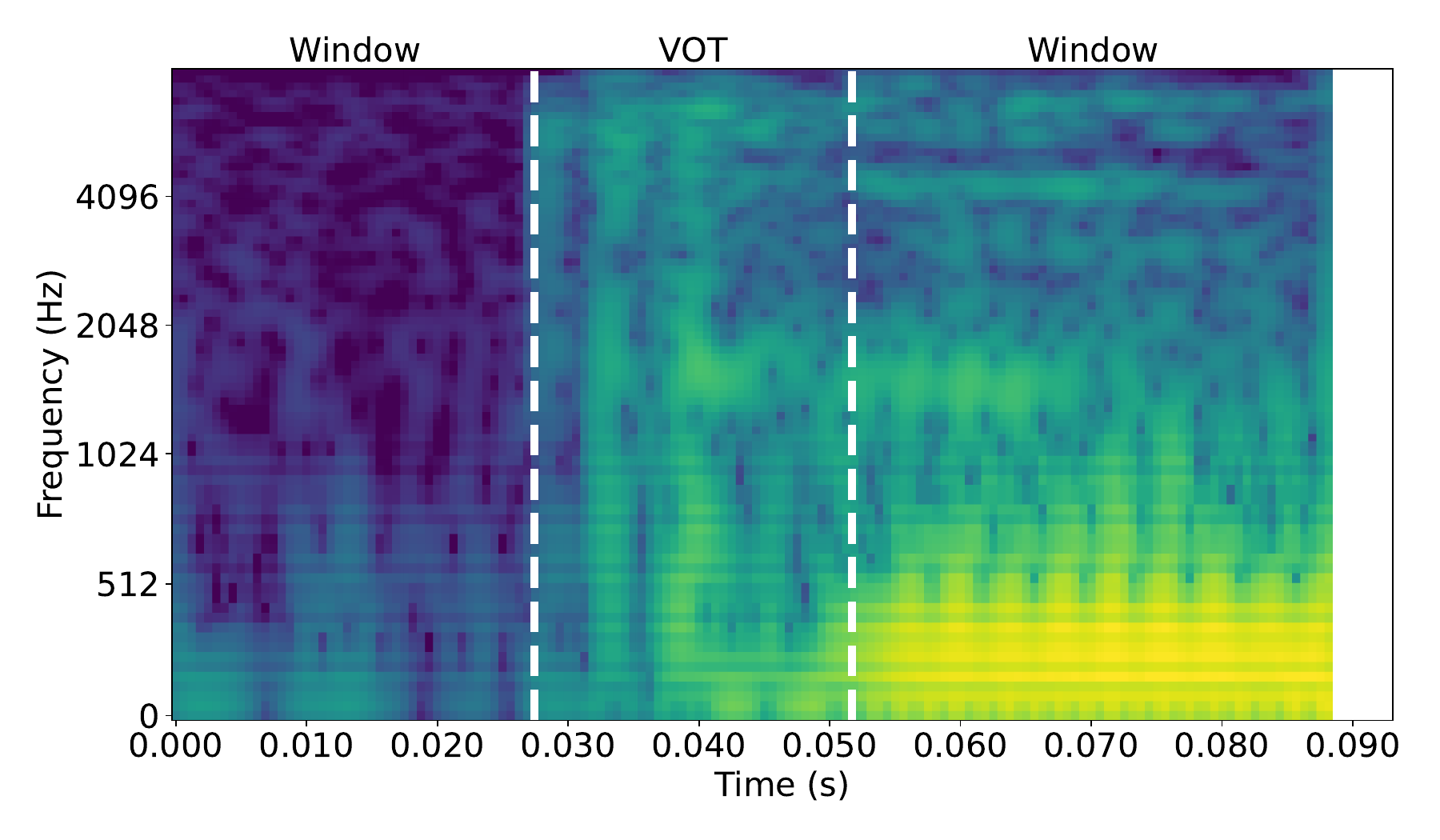}
    \includegraphics[width=0.33\linewidth]{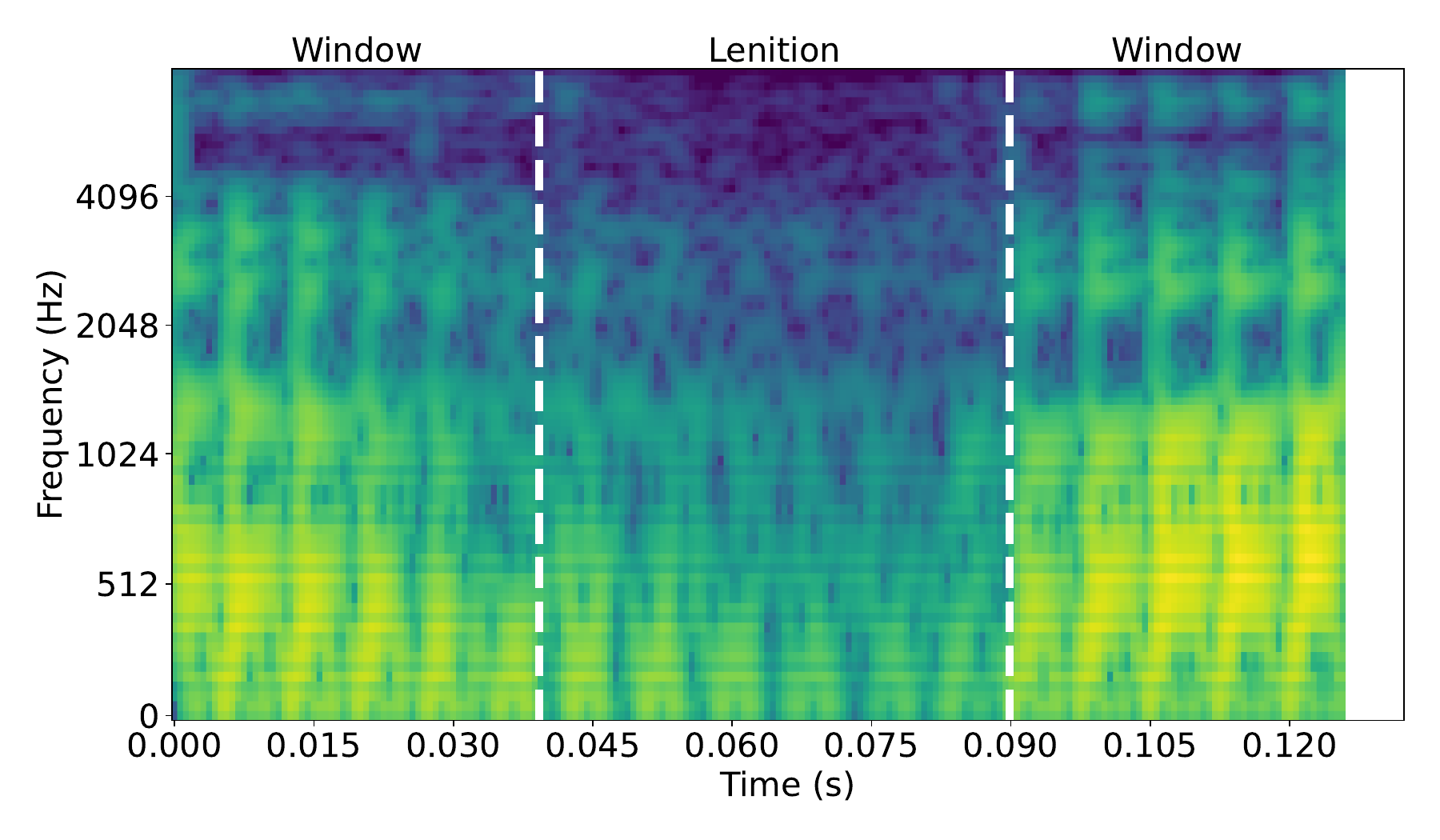}
    \caption{Spectograms of three stop tokens from the Corpus of Spontaneous Japanese-Core (CSJ-C), overlaid with Closure + VOT (left), VOT (middle), and lenition (right) predicted from wav2VOT model trained in Section \ref{subsubsec:csj}.}
    \label{fig:predspec}
\end{figure*}

The architecture of wav2vec2 consists of a Feature Encoder (FE) block and Transformer Encoder (TE) block. Feature encoding in wav2vec2 is performed via a set of 1-dimensional convolutional layers which reduce the dimensionality of the speech signal to a set of features based on the product of the strides of all convolutional layers. In the standard wav2vec2 configuration, the FE consists of 7 1-D convolutional blocks with strides of (5, 2, 2, 2, 2, 2, 2), resulting in an effective time resolution of 1 frame per 20ms.\footnote{A set of convolutional layers with strides of (5, 2, 2, 2, 2, 2, 2) downsamples the audio vector by a factor of 320; 1 second of audio (at a sampling rate of 16kHz) is downsampled to 50 frames (16000 /\ 320 = 50), each constituting 20ms of audio.} The output of the FE is subsequently passed to the TE block, which outputs a set of hidden states corresponding to each frame output by the FE. For use in a `downstream' task (such as classification or regression), a head layer is placed after the final state of the TE, which may either output a single value or multiple frame-wise values. Wav2VOT performs such classification on frame-wise basis by predicting one of 4 labels \{surrounding context window, closure, VOT, lenition\} for each output frame. As the temporal transitions between properties such as VOT and closure duration are more fine-grained than the default 20ms temporal resolution of wav2vec2, wav2VOT instead uses 4 convolutional layers with a stride of (2, 2, 2, 2), resulting in an downsampling factor of 16 and an effective time resolution of 1ms.

For training, stop tokens are extracted with a surrounding context window\footnote{To avoid overfitting to the size of the window (i.e. learning that closure onset is e.g. 30ms from the start of the token), the length of the start and end windows are randomly drawn from a uniform distribution with user-defined upper and lower bounds.} along with metadata delineating the closure onset, VOT start, VOT end, and optionally whether the stop token has undergone reduction. These timepoints are then used to generate a vector sequence of labels corresponding to each 1ms of the extracted stop token (e.g. [0, 0, 0, 0, 1, 1, 1, 1, 2, 2, 2, 2, 2, 0, 0, 0, 0], where 0 = window, 1 = closure, 2 = VOT). To stabilise the training process and ensure the model learns the set of possible sequences (e.g. that closures cannot follow VOT), a Connectionist Temporal Classification (CTC) loss is calculated on the collapsed predicted sequence (e.g. [0, 0, 1, 1, 1, 2, 2, 0, 0, 0] $\rightarrow$ [0 1 2 0]), which is scaled by 0.05 and added to the framewise cross-entropy loss on the labels.

For inference, wav2VOT returns a sequence of frame-wise softmax-predicted labels (e.g. [0, 0, 1, 1, 1, 2, 2, 0, 0, 0]). When these frame-wise predictions are converted to label start/end timestamps, to prevent the generation of frame-length intervals, a minimum interval size (default 5ms) is enforced, where prospective intervals shorter than the minimum are combined with the previous label. Figure \ref{fig:predspec} illustrates wav2VOT predictions overlaid on examples of a (word-medial) combined closure duration + VOT sequence, a (word-initial) VOT-only sequence, and an example of a lenited token with no distinct burst.

% -- ie outputs 1 frame per 20ms of audio.

% - as the transitions between CD and VOT are more temporally fine-grained than 20 ms, the structure of the FE was changed in wav2vot
% - instead of 7 1D convolutional layers, wav2vot instead uses 4 convolutional layers with a stride of (2, 2, 2, 2), resulting in an downsampling factor of 16 and an effective time resolution of 1ms\footnote{Add something here about how this changes the receptive field}
% - wav2vot thus outputs 1ms-long frame-wise predictions for the closure and VOT of a stop (make figure that shows framewise preds on top of a spectrogram)

    % - note somewhere that these are logits that are softmaxed
% - to avoid issues of rapid changes in the prediction within a stop (e.g. 1ms-wise flucations between closure and VOT), two additional steps:
% - 1. during training, a CTC loss is calculated on the stop 'sequence' (e.g. WINDOW CLOSURE VOT WINDOW); this CTC loss (scaled by 0.05) is combined with the framewise cross-entropy loss to regularise for learning the sequence of labels (as well as the framewise labels)

% - 2. when framewise predictions are converted to intervals (following inference), a minimum interval size (default 5ms) is enforced, where prospective intervals shorter than the minimum are combined with the previous label

\section{Experiment 1}
\label{sec:ft}
The goals of Experiment 1 are to demonstrate that wav2VOT is capable of learning and capturing the fine-grained temporal properties of stops in 1) a large set of training data with a matched test set (Sec.~\ref{subsubsec:csj}), and 2) across a variety of unseen datasets, along with changes in performance when fine-tuned on varying amounts of data (Sec.~\ref{subsubsec:ft}).

% explore the predictive performance wav2VOT. This is performed by first training a wav2VOT model on a large amount of manually-annotated stops and evaluate its performance on predicting from stops drawn from the same data . This is then extended to the prediction of stops from unseen datasets and the impact on performance when wav2VOT is a drawn from those datasets.

% structure of experiment:
% - first show initial training on large speech corpus with VOT/closure (and reduction) annotations (CSJ)
% - show predictive performance on selection of other speech datasets, including performance changes with differing amounts of finetuning data

\subsection{Initial Training}
\label{subsubsec:csj}

The data for the initial model training comes from the `Core' section of the \emph{Corpus of Spontaneous Japanese} (CSJ-C) \cite{Maekawa2000,koiso2014}, a large speech corpus containing 45 hours of spontaneous Japanese monologues from 139 speakers. Japanese employs a two-way stop system, where voiced stops are realised with short VOT and closure duration, and voiceless stops with intermediate VOT and long closures \cite{shimizu1996,riney2007}. The CSJ-C contains extensive manually-corrected segmental and phonetic annotation, including separate closure and VOT annotations for stops, as well as distinct annotations for instances where a separate burst could not be observed in the speech signal (i.e. reduced).

\begin{figure}[h]
    \centering
    \includegraphics[width=0.7\linewidth]{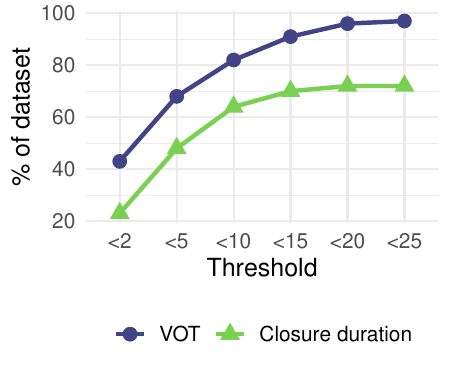}
    \caption{Performance of CSJ-C-trained wav2VOT model for the CSJ-C training set, reported as a percentage of the test corpora that fall within a set of fixed tolerances (e.g., $\textless$2ms refers to a proportion of predictions within 2ms of the manual annotations).}
    \label{fig:csj_pred_plot}
\end{figure}

For training, audio for 205,034 stop tokens was extracted from the CSJ-C, with the size of the surrounding context window for each stop token drawn from a uniform distribution with limits of 30-60ms. A wav2VOT model was initialised with random weights and trained for 10 epochs (batch size = 64) using a 80GB NVIDIA H100 GPU. The model was evaluated for framewise accuracy, framewise F1, and sequence word-error-rate (WER) on a randomly-drawn 41,000-token (20\%) test set every 200 steps. The best-performing model achieved an evaluation performance of \{framewise accuracy = $96.4\%$, framewise F1 = $0.93$, WER = $0.06$\}. Following the practice of evaluating AutoVOT performance \cite{autovot}, we evaluated the performance of this wav2VOT model as a proportion of the validation set predictions below a set of thresholds (e.g., $\textless$2ms, $\textless$5ms, etc). The VOT and closure duration performance is summarised in Figure \ref{fig:csj_pred_plot}, which demonstrates that the majority of VOT predictions are within 5ms of error; closure duration is estimated with less accuracy for the CSJ-C, reflecting the relative difficulty to closure duration annotation compared with VOT \cite{sondereggerlanguage}. Burst realisation was predicted by this model with $93.3\%$ accuracy (F1 = $0.96$), and the stop sequence (i.e., closure + VOT vs VOT) was predicted with $96.8\%$ accuracy (F1 = $0.8$).

% - explain CSJ structure and annotation scheme
% - explain data preparation:
%     - audio extraction
%     - conversion to framewise labels

% - training process:
%     - hardware, epochs, saving process etc

% - report eval metrics

\subsection{Finetuning Experiments}
\label{subsubsec:ft}

\begin{figure*}[h]
    \centering
    \includegraphics[width=0.94\linewidth]{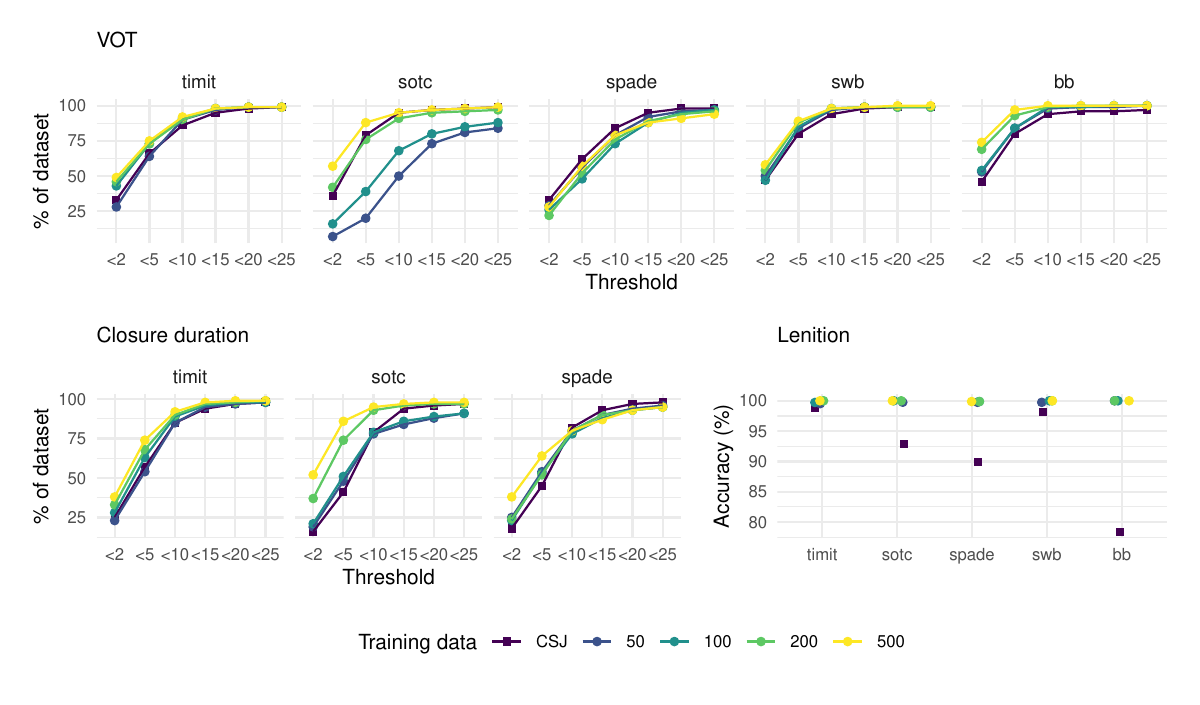}
    \caption{Performance of wav2VOT VOT (top) and closure duration (bottom left) predictions, reported as a percentage of the test corpora that fall within a set of fixed tolerances. Lenition predictions (bottom right) reported as predictive accuracy (with horizontal jitter). Purple squares lines indicate the CSJ-C-trained model (i.e. no finetuning), with circles corresponding to models finetuned with different amounts of training data from the given corpus.}
    \label{fig:ft_pred_plot}
\end{figure*}

Having demonstrated that wav2VOT can predict VOT, closure duration, and burst realisation in a large matched test set, the question then follows whether this performance can be observed when applied to other speech datasets. Usage of AutoVOT recommends that users train a classifier model on a subset of the target data, such that the parameters reflect the target distribution: in order to determine whether this recommendation should also apply for wav2VOT, it is necessary to test whether the fine-tuning of a trained wav2VOT model with data from a given speech corpus results in improved predictive accuracy. Specifically, this section evaluates the quality of VOT, closure duration, and lenition predictions with a set of English speech corpora, comparing the performance of the baseline model trained only with CSJ-C data (as reported in Sec.~\ref{subsubsec:csj}) with models fine-tuned on increasing amounts of training data drawn from that corpus.

% having demonstrated that wav2vot can be trained to predict vot/closures/reduction, main real-world usecase is how this can be applied to new speech unseen speech data
% shown with autovot: performance with the OOB model is good, but recommendation is to train a new classifier on data from the target distribution (i.e. from the same dataset)
    % - this is common view in phon research (i.e. 'overfitting' to the target dist) since accuracy is preferred over generalisability (see also eg MFA); of course, like e.g. MFA, goal would be to have a system that is accurate but can also generalise to new cases (i.e. to make usage easier)
% thus focus here is to eval the performance of wav2vot on stops from other speech corpora, testing:
% 1. the performance of the CSJ-only model;
% 2. whether finetuning wav2vot on varying amounts of data from target distribution improves performance

\subsubsection{Method}

The five speech corpora selected for this experiment were chosen based on a number of factors, including variability in English dialects, speech styles, and recording contexts across corpora. The five selected corpora were as follows:
\begin{itemize}
    \item \emph{TIMIT} \cite{timit}: 14389 word-medial stop tokens from read sentences by 630 American English speakers, annotated for closures and VOT.
    \item \emph{Sounds of the City} (SOTC \cite{sotc}): 12555 stop tokens from spontaneous sociolinguistic interviews with 23 Glasgow vernacular English speakers, manually annotated for VOT, closure duration, and lenition as reported in \cite{sondereggerlanguage}.
    \item \emph{SPeech Across Dialects of English} (SPADE \cite{dmuc}): 1903 stop tokens from a meta-corpus of 17 British and North American English dialects, consisting of a combination of read and spontaneous speech styles, manually annotated for VOT, closures, and lenition by the first author.
    \item \emph{Switchboard} (SWB \cite{switchboard}): 893 word-initial stop tokens from 16 American English speakers, taken from spontaneous telephone speech. Annotated for VOT as reported in \cite{autovot}.
    \item \emph{Big Brother} (BB \cite{sonderegger2017}): 704 word-initial stops tokens from 4 British English speakers recorded in the 2008 season of Big Brother UK, manually annotated for VOT as reported in \cite{autovot}.
\end{itemize}

The same fine-tuning and evaluation procedure was applied to each corpus. For each corpus, the stop audio was extracted from the corpus based on the provided annotations (as in Sec.~\ref{subsubsec:csj}). A training `master' set of 500 tokens was selected from that corpus, with $N$-500 tokens used as that corpus's validation set. From this 500-stop training set, a wav2VOT model was fine-tuned with random subsets, corresponding to 50, 100, 200, and 500 tokens for 10 epochs. VOT, closure duration, and lenition were subsequently predicted for the validation set.

% - explain finetuning procedure
%     1. process training data as above (i.e. extract stop audio plus timing metadata)
%     2. take 500 stops as a subset
%     3. use 500-stop subset finetune models based on 50, 100, 200, and 500
%     4. predict the holdout (full - 500) using CSJ and those models
%     5. following autovot, calculate the perc of holdout data that is within a threshold from the annotated VOT

\subsubsection{Results}

The performance of the fine-tuned models on each of the five corpora are shown in Figure \ref{fig:ft_pred_plot}. Looking first at the ability of wav2VOT to predict for unseen data, we observe that the CSJ-C-only model (purple squares) performs well in estimating VOT in these cases (Fig.~\ref{fig:ft_pred_plot}, top). While wav2VOT reports a smaller proportion of stops estimated within 2ms than reported for AutoVOT \cite{autovot} (e.g. 47\% vs AutoVOT's 53\% for SWB; 46\% vs 53\% for BB), wav2VOT reports a greater proportion of stops within the 5ms threshold (80\% vs 73\% for SWB; 80\% vs 79\% for BB), suggesting that wav2VOT performs comparatively with AutoVOT with respect to estimating from unseen data. This pattern holds also for closure duration (Fig.~\ref{fig:ft_pred_plot}, bottom left), where the majority of closures are estimated within 10ms, though a smaller proportion of closures are estimated with 2ms error than VOT. The CSJ-C-only model shows variable performance for lenition prediction (Fig.~\ref{fig:ft_pred_plot}, bottom right), though accuracy $\geq$90\% is observed for all corpora except for BB, comparable with similar tools for predicting burst realisation \cite{tanner2025interspeech}.

Next, looking at the effect of increased finetuning data, we observe that, for the two datasets with most variability in recording quality (SOTC and SPADE), small amounts of data (50-100 samples) in finetuning appears to damage performance on VOT and closure duration relative to the CSJ-C-only model. For the remaining corpora, however, finetuning wav2VOT using a moderate amount of samples (200+) results in improved VOT and closure duration accuracy (i.e. a greater proportion of stops estimated within 2-5ms of the manual annotations). Finetuning improves the prediction of lenition for all corpora, suggesting that allowing wav2VOT to learn the data-specific speech style and recording context results in higher-quality predictions. 

\section{Experiment 2}
\label{sec:analysis}

Following previous work evaluating the performance of automatic VOT annotations \cite{autovot,buz2018}, we also evaluate the predictive performance of wav2VOT by comparing predictions directly with manual annotations in a hypothetical phonetic study context. Specifically, we take a regression modelling approach to estimate the degree of similarity between wav2VOT predictions and manual annotations while accounting for other sources of variability known to affect VOT and closure duration.

Using the TIMIT validation set and predictions from the model finetuned on 500 TIMIT stops (as reported in Sec~\ref{subsubsec:ft}; Fig.~\ref{fig:ft_pred_plot}, top), VOT and closure duration were jointly modelled in a Bayesian regression model \cite{brms}. The source of the value (\{annotation, wav2VOT\}), the voicing category of the stop, place of articulation (POA), and speech rate were included as predictors, along with all two-way interactions with annotation source, and the three-way interaction with source, voicing, and POA. Speaker-level random intercepts and slopes were also included for all fixed-effect predictors, along with the speaker-level relationship between VOT and closure duration intercepts.

% As VOT and closure duration follow normal distributions in log-space, both responses were fit with the $lognormal$ family, with $normal(3.5,0.5)$ and $normal(4,0.5)$ priors for the VOT and closure duration intercepts respectively, $normal(0,1)$ for the fixed-effect ($\hat{\beta}$) terms, $exponential(1)$ for the random-effect ($\hat{\sigma}$) terms, and $lkj(1.5)$ for the between-response correlation ($\hat{\rho}$) term. The posterior distribution was sampled for 2000 (1000 warmup) samples per chain (4 chains) using the \emph{cmdstanr} backend \cite{cmdstanr}.

% - compare how wav2vot predictions compare with annotations in e.g hypothetical study
% - fit baysian model of VOT + CD of TIMIT stops
% - predictors:
%     - source of value (annotation vs w2v)
%     - voicing
%     - poa
%     - SR
%     - interactions (incl method-voicing-poa)
%     - speaker ranefs (int, method, voicing corred across vars; poa + sr)

\subsection{Results}
\begin{figure}[h]
    \centering
    \includegraphics[width=\linewidth]{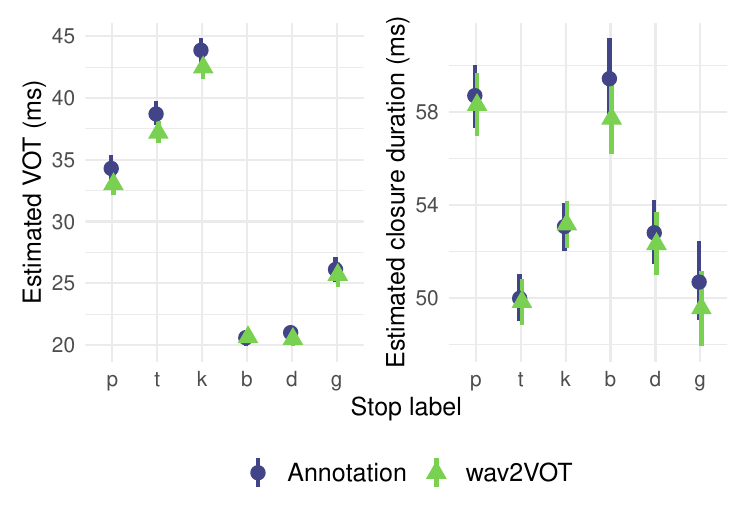}
    \caption{Model-estimated VOT (left) and closure duration (right) for manually-annotated stops (purple circles) and wav2VOT predictions (green triangles), separated by phoneme label. Points indicate posterior median, with lines indicating the 95\% Highest Posterior Density (HPD) interval \cite{emmeans2024}.}
    \label{fig:vot_pred_plot}
\end{figure}

% In this experiment, the predictive performance of wav2VOT is evaluated based on the similarity between estimated values for manual annotations and wav2VOT predictions as estimated from a regression model, focusing specifically on the overall estimated difference, as well as the difference between properties of stops well-known to differ: phonological voicing category and place of articulation. Results are reported as summaries of the model's posterior distribution, including the median and 95\% Credible Interval (CrI) of a given model term, as well as the posterior probability of a hypothesis (i.e. the probability of a parameter's distribution containing zero).

% \begin{figure}[h]
%     \centering
%     \includegraphics[width=0.7\linewidth]{cd_pred_plot.pdf}
%     \caption{Model-estimated closure duration values for manually-annotated stops (purple circles) and wav2VOT predictions (green triangles), separated by phoneme label. Points indicate posterior median, with lines indicating the 95\% Highest Posterior Density (HPD) interval \cite{emmeans2024}.}
%     \label{fig:cd_pred_plot}
% \end{figure}

Figure \ref{fig:vot_pred_plot} (left) shows the model estimated VOT for manual annotations and wav2VOT predictions. Turning first to VOT, we do not observe evidence of an overall difference in VOT between manual annotations and wav2VOT predictions. While methods differ in the size of the estimated voicing contrast, the size of this difference is negligible, corresponding to a contrast of 10.76ms for manual annotations and 10.5ms for wav2VOT.

As with VOT, we observe neglible differences between manual annotations and wav2VOT in closure duration (Fig.\ref{fig:vot_pred_plot}, right). No difference in closure duration as a function of phonological voicing is observed; in turn, there is no evidence for a difference between methods in the voicing contrast. We also do not observe any differences across places of articulation.\footnote{The full model table is reported in the Supplementary Materials.}

% ($\hat{\beta}_{method} = -0.08, CrI = [-0.27,0.1], P(\hat{\beta}<0)=0.8$)
% ($\hat{\beta}_{method\times voicing} = 0.02, CrI = [0,0.05], P(\hat{\beta}>0)=0.96$)
% ($\hat{\beta}_{method\times POA1} = -0.01, CrI = [-0.02,0.01], P(\hat{\beta}<0)=0.77$; $\hat{\beta}_{method\times POA2} = 0, CrI = [-0.01,0.01], P(\hat{\beta}<0)=0.43$)

% - find that little/no evidence for difference between annotation and w2v
% - as shown in plot, w2v captures the voicing/poa-specific VOTs

% ($\hat{\beta}_{method} = -0.01, CrI = [-0.16,0.14], P(\hat{\beta}<0)=0.53$)
% ($\hat{\beta}_{voicing} = 0.09, CrI = [-0.18,0.35], P(\hat{\beta}>0)=0.74$)
% ($\hat{\beta}_{method\times voicing} = -0.02, CrI = [-0.04,0], P(\hat{\beta}<0)=0.95$)
 % ($\hat{\beta}_{method\times POA1} = 0, CrI = [-0.01,0.02], P(\hat{\beta}>0)=0.8$; $\hat{\beta}_{method\times POA2} = 0, CrI = [-0.01,0.01], P(\hat{\beta}<0)=0.61$)

% - same as VOT above -- no meaningful difference between w2v-predicted CD and those from the annotations
% - also follows stop-specific CD patterns

\section{Discussion}
\label{sec:discussion}

The goal of this study has been to further explore the application of large speech models to the task of phonetic annotation. In particular, we sought to explore the extent to which wav2vec2 can accurately estimate the acoustic properties of stops, and introduce wav2VOT: an application of wav2vec2 for the automatic estimation of VOT, closure duration, and burst realisation. In Experiment 1 (Sec.~\ref{sec:ft}), we observe that, without prior finetuning, wav2VOT performs comparably with other tools for predicting properties of stops \cite{autovot,tanner2025interspeech}, and finetuning for a given target dataset further improves predictive accuracy. In Experiment 2 (Sec.~\ref{sec:analysis}), we observe negligible quantitative differences between approaches in overall VOT and closure duration, as well as across phonological voicing or place of articulation.

% - goal of this study is to further explore the application of large speech models to the task of phonetic annotation

% - specifically here explore application of wav2vec2 for estimating the temporal properties of stop segments -- chiefly VOT and CD, which are well-studied acoustic parameters that vary as a function of linguistic, phonological, and speaker diffs: in this vain report wav2VOT: an application of wav2vec2 for VOT and closure duration estimation

% - across 2 experiments:
    % - 1. demonstrate that wav2VOT has comparable performance to AutoVOT (widely used tool for VOT estimation) on unseen datasets that vary substantially in dialect, speech style, and recording context, and can be fine-tuned for high accuracy on a given target dataset
    % - 2. show that predictions from wav2VOT closely follow manual VOT and closure duration annotations similar to a hypothetical phonetic study setup

The results of this study demonstrate that the properties of stops can be accurately estimated with the wav2vec2 architecture, and provide further evidence for the potential of speech models for performing speech annotation tasks. As wav2vec2 is designed without a pre-defined use-case, this versatility makes it powerful for a wide range of speech annotation and segmentation tasks \cite{kim2024,tanner2025interspeech,rousso2024traditioninnovationcomparisonmodern}. This, in turn, motivates future work exploring the capacity of wav2vec2 and similar models \cite{wavlm,whisper2022} for performing phonetic annotation (including phonetic properties of stops such as closure voicing, pre-aspiration, and negative VOT), along with validation across the linguistic and stylistic variability common in phonetic research, including differing three-way systems and implementations of stop realisation.

% - this study provides further evidence for the potential of large speech models for application to annotation tasks common in phonetic research pipelines
    % - the relative 'blank-slate' of the wav2vec2 architecture (which does not have a pre-defined downstream usecase) is particularly powerful in this sense, as this makes it versatile for a wide range of different annotation and speech processing tasks.
    % - basically: wav2vec2 can be modified to predict pretty much any kind of acoustic variability with (relative) ease, given some training dataset exists

% - this in turn motivates future work that explores the capacity of wav2vec2 and similar speech models for the annotation of speech data for linguistic research, which has the potential to lower the financial and resource barriers to performing phon research on large and diverse datasets

% Uncomment for final version
\section{Acknowledgements}
The authors thank the SPADE Data Guardians, Rachel Macdonald, Michael McAuliffe, and Vanna Willerton. Computational resources were provided by the Digital Research Alliance of Canada. This research was supported by a T-AP Digging into Data award in the form of the following grants: ESRC Grant \#ES/R003963/1, NSERC/CRSNG Grants \#RGPDD-501771-16 and \#RGPIN-2023-04873, SSHRC/CRSH Grant \#869-2016-0006, and NSF Grant \#SMA-1730479. This research was also supported by a Canada Research Chair \#CRC-2023-00009 (MS), and by a British Academy Postdoctoral Fellowship, a University of Glasgow Lord Kelvin/Adam Smith Fellowship, and the University of Glasgow's Centre for Data Science \& AI (JT).

\section{Generative AI Use Disclosure}
The authors declare that no Generative AI tools were used in the research, design, analysis, writing, or editing of this work. 

\bibliographystyle{IEEEtran}
\bibliography{JTanner}
\end{document}